\documentstyle[twocolumn,aps,prl,floats,epsf]{revtex}

\newcommand{\pbar}{\overline{p}}

\newcommand{\ks}{K^0_{s}}

\newcommand{\qbar}{\overline{q}}
\newcommand{\ppair}{\pbar p}

\newcommand{\ksks}{\ks\ks}

\newcommand{\pbarp}{\ppair}

\newcommand{\qqbar}{q\qbar}

\newcommand{\phiphi}{\phi\phi}
\newcommand{\pptopppipi}{\pbarp\rightarrow\pbarp\pi^{+}\pi^{-}}

\newcommand{\pptofourk}{\pbarp\rightarrow 4K^{\pm}}

\newcommand{\pptophiphi}{\pbarp\rightarrow \phiphi}
\newcommand{\pptophikk}{\pbarp\rightarrow \phi K^{+}K^{-}}

\newcommand{\brxtopbarp}{\rm{B}(\xi\rightarrow\pbarp)}

\newcommand{\brxtoksks}{\rm{B}(\xi\rightarrow\ksks)}

\newcommand{\lt}{<}

\newcommand{\degree}{\mbox{$^{\circ}$}}
\newcommand{\etal}{{\em et al.}}

\begin{document}

\wideabs{

\title{Study of the reaction $\bar p p \to \phi \phi$ from 1.1 to
       2.0 GeV/$c$}

\author{C.~Evangelista and A.~Palano}
\address{University of Bari and INFN, I-70126 Bari, Italy}

\author{D.~Drijard, N.H.~Hamann\cite{by1}, R.T.~Jones\cite{by2}, B.~Mou\"{e}llic, S.~Ohlsson and J.-M.~Perreau}
\address{CERN, European Organization for Nuclear Research, CH-1211 Geneva, Switzerland}

\author{W.~Eyrich, M.~Moosburger, S.~Pomp and F.~Stinzing}
\address{University of Erlangen-N\"{u}rnberg, W-8520 Erlangen, Germany}

\author{H.~Fischer, J.~Franz, E.~R\"{o}ssle, H.~Schmitt and H.~Wirth\cite{by3}}
\address{University of Freiburg, W-7800 Freiburg, Germany}

\author{A.~Buzzo, K.~Kirsebom, M.~Lo Vetere, M.~Macr\`{\i}, M.~Marinelli, S.~Passaggio, M.G.~Pia, A.~Pozzo, E.~Robutti and A.~Santroni}
\address{University of Genova and INFN, I-16132 Genova, Italy}

\author{P.T.~Debevec, R.A.~Eisenstein\cite{by4}, P.G.~Harris\cite{by5}, D.W.~Hertzog, S.A.~Hughes, P.E.~Reimer\cite{by6} and J.~Ritter}
\address{University of Illinois, Urbana, Illinois, 61801 USA}

\author{R.~Geyer, K.~Kilian, W.~Oelert, K.~R\"{o}hrich\cite{by3}, M.~Rook and O.~Steinkamp}
\address{Institut f\"{u}r Kernphysik, Forschungszentrum J\"{u}lich, 
D-5170 J\"{u}lich, Germany}

\author{H.~Korsmo and B.~Stugu\cite{by7}}
\address{University of Oslo, N-0316 Oslo, Norway}

\author{T.~Johansson}
\address{Uppsala University, S-75121 Uppsala, Sweden}

\author{(The JETSET Collaboration)}

\date{18 February 1998}

\maketitle

\begin{abstract}
A study has been performed of the reaction $\pptofourk$ using
in-flight antiprotons from 1.1 to 2.0 GeV/$c$ incident momentum
interacting with a hydrogen jet target.  The reaction is dominated by
the production of a pair of $\phi$ mesons.  The $\bar p p \to \phi
\phi$ cross section rises sharply above threshold and then falls
continuously as a function of increasing antiproton momentum.  The
overall magnitude of the cross section exceeds expectations from a
simple application of the OZI rule by two orders of magnitude.  In a
fine scan around the $\xi/f_J(2230)$ resonance, no structure is
observed.  A limit is set for the double branching ratio
B($\xi\rightarrow\pbarp)\times\rm{B}(\xi\rightarrow\phi\phi) \lt 6
\times 10^{-5}$ for a spin 2 resonance of M = 2.235 GeV and $\Gamma$ =
15 MeV.
\end{abstract}

} 

\section{Introduction}

Beyond the common families of baryonic ($qqq$) and mesonic $(q \bar
q)$ states, glueballs ($gg$ or $ggg$), hybrids ($q \bar q g$) and
multi-quark systems ($e.g., q \bar q q \bar q$) are predicted to exist
based on the current understanding of QCD and QCD-inspired
phenomenological models of hadrons.  The question of the existence of
these unusual hadronic bound systems is fundamental and should lead to
a broader understanding of the behavior of QCD in the non-perturbative
region.  The experiment described below investigates the existence of
gluonic hadrons in the mass range between 2.1 and 2.4 GeV/$c^{2}$.
The technique requires a state, if found, to couple appreciably to
both antiproton-proton (the entrance channel) and $\phi\phi$ (the exit
channel).  The parameter space explored is of particular interest in a
search for the tensor glueball.

Gluonic hadrons are expected to populate the low mass region of the
hadron spectrum together with normal $q \bar q$ states.  Recent
lattice QCD calculations locate the scalar ($J^{PC}=0^{++}$) glueball
mass in the range of 1.5 to 1.7 GeV~\cite {latty}.  The tensor
($J^{PC}=2^{++}$) is the next low-lying glueball and is found to be
approximately 1.5 times more massive than the scalar, thus it is
postulated that it might be in the region explored by this experiment.

Several experimental approaches have been used to search for
glueballs.  The common theme is to choose a process where a
kinematical or dynamical filter strongly reduces the large production
of standard $q \bar q$ mesons.  Reactions intensively explored are
peripheral hadron interactions ~\cite {chung}, $J/\psi$ radiative
decays ~\cite {wermes}, central production ~\cite {sandiego}, $\gamma
\gamma$ collisions~\cite {feindt} and $\bar p p$ annihilations ~\cite
{amsler}.

The study of the $\pptophiphi$ reaction is motivated by the
expectation that, with essentially no common valence quarks between
the initial and final states, the reaction is forbidden in first
order.  Indeed, a strict application of the empirical
Okubo--Zweig--Iizuka (OZI) rule~\cite{ozi} suggests a cross section
for $\pptophiphi$ of approximately 10 nb.  In a previous
publication~\cite{berto}, we reported the cross section at 1.4 GeV/$c$
incident antiproton momentum to be $3.7 \pm 0.8$~$\mu$b and thus two
orders of magnitude greater than estimated.  The large $\pptophiphi$
cross section has been interpreted as coming from two-step
processes~\cite{lipkin,mull}, from intermediate glue~\cite{linde},
and/or from the intrinsic strangeness in the nucleons~\cite{ellis}.

Here we review the appropriateness of the energy range of our study
with respect to the possibility of exciting a gluonic resonance.  As
evidence, consider the current picture of candidate glueball states.
The $f_0(1500)$, discovered by the Crystal Barrel experiment at the
CERN Low-Energy Antiproton Ring (LEAR), has a width of $\Gamma\approx$
120 MeV and has been observed to decay to $\pi^0 \pi^0, \eta
\eta$~\cite{cbarr}, $\eta \eta'$~\cite{cbarr1} and $K \bar
K$~\cite{cbarr2} following $\ppair$ annihilation at rest. The
existence of this state has been confirmed in in-flight $\bar p p$
annihilations by experiment E760 at Fermilab ~\cite{e760}.  It has
been interpreted as the leading scalar glueball candidate.  Alternate
explanations feature strong gluonic content; the state could be a
mixture of a glueball and nearby ordinary $\qqbar$ systems~\cite
{close}.  If identified with the scalar glueball, the implication from
the lattice calculations is that the tensor mass should be in the 2.1
to 2.4 GeV range where candidate glueball states have already been
identified.  These states include the $\xi/f_J(2230)$ which is
observed in radiative $J/\psi$ decays to $K \bar K$ ~\cite{mkIII} and
also in decays to $\pi \pi$ and $p \bar p$~\cite{bes}.  The width of
this state ($\Gamma \approx 30 \pm 20$ MeV) is unexpectedly small, and
has stimulated considerable interest. The spin-parity is undetermined
but is limited to (even)$^{+}$.  This channel is of interest to the
present experiment since it couples to $\ppair$.  In another
study~\cite{jetset-2} from this experiment, stringent limits were
established on the double branching ratio $\brxtopbarp \times
\brxtoksks$ for the case of a narrow (i.e., $\Gamma < 30$ MeV) state.

Resonances in the $\phi\phi$ system have been observed in
$J/\psi\rightarrow\gamma\phi\phi$ by Mark~III and DM2.  Each group saw
a wide bump above $\phi\phi$ threshold peaking near 2.2 GeV and having
a width of approximately 100 MeV.  The preferred $J^P$ assignment is
$0^-$, however $2^+$ contributions cannot be excluded
~\cite{phiphijpsi}.

The $\phi\phi$ system has been studied in exclusive $\pi^- p$
~\cite{arm,linde1} and inclusive $\pi^-Be$ induced
interactions~\cite{booth}.  This reaction is expected to be highly
suppressed based on the above OZI argument related to disconnected
quark-line diagrams.  However the suppression is not observed near
threshold, a fact which could be explained by the production of
intermediate gluonic resonances~\cite{linde}. A partial wave analysis
(PWA) of the exclusive data revealed the presence of three interfering
$J^{PC}=2^{++}$ resonances, the $f_{2}(2010), f_{2}(2300)$ and
$f_{2}(2340)$, originally called $g_T$ states.  The inclusive data
show evidence for two structures with parameters compatible with those
measured for the $g_T$ states but a PWA was not possible.

We report below on the evaluation of 58 independent measurements of
the reaction $\pptofourk$ from which we determine the cross sections
for $\pptophiphi$, $\pptophikk$ and $\pptofourk$.  A threshold
enhancement is evident in the $\phi\phi$ cross section data.  In the
narrow region around the $\xi/f_J(2230)$ resonance, 17 measurements of
the $\phi\phi$ cross section are reported in narrow energy steps
covering a range of 45 MeV.  Limits are set on the non-observation of
this state.

\section{The experiment}

The JETSET (PS202)~\cite{nikko} experiment is designed to measure the
reaction $$\bar p p \to \phi \phi \to K^+ K^- K^+ K^- \eqno(1)$$ as a
function of incident antiproton momentum from 1.1 GeV/$c$ to the
highest momenta available at LEAR (2.0 GeV/$c$).  The experiment makes
use of the LEAR antiproton beam incident on an internal target
provided by a hydrogen-cluster jet system.  To obtain a measure of the
absolute (and relative) luminosity at different antiproton momenta,
$\pbarp$ elastic scattering events were recorded in parallel using a
special dedicated trigger.

One key feature of reaction (1) in the energy range covered by this
experiment is the fact that the outgoing kaons are constrained to a
forward cone below $60\degree$.  Dominant in $\pbarp$ annihilation are
the production of charged and neutral pions.  These unwanted
background reactions produce relatively fast charged particles and
photons spread over a much larger angular range.  The design of the
detector and electronic trigger takes advantage of these facts by
providing good charged-particle tracking and particle identification
(PID) for forward angles, fast charged-particle multiplicity for
triggering, and a large-acceptance photon rejection system.  The
philosophy followed to record and identify $4K^{\pm}$ events is to
trigger only on the proper multiplicity pattern.  In the offline
analysis, events are required to have four well-determined tracks in
the right kinematical range.  Finally, these tracks are associated
with hits in the PID detectors to determine compatibility with
identification as kaons from the $4K$ final state.

\begin{figure}
  \begin{center}
    \mbox{\epsffile{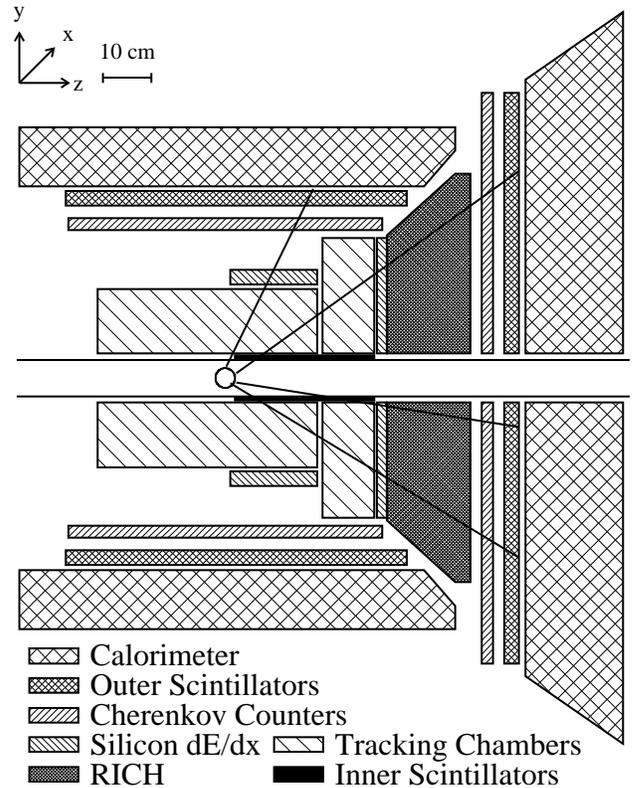}}
  \end{center}

  \caption{ Layout of the JETSET/PS202 experiment, showing the major
  detector components.  A typical event with three forward and one
  barrel track is shown.}

\label{fig0}
\end{figure}

A schematic view of the apparatus is shown in Fig.~\ref{fig0}.  Its
basic structure is a hydrogen-cluster jet target (not
shown)~\cite{Mario} with a density $\rho = 4 \times
10^{12}$~atoms/cm$^2$, surrounded by a compact detector.  The jet is
oriented horizontally and intersects the circulating antiproton beam
at right angles.  Around the interaction volume the LEAR ring is
equipped with an oval vacuum chamber (0.03 radiation lengths thick)
having horizontal and vertical half-axes of 78 and 38~mm,
respectively.  This limits the geometrical acceptance of the detector
to polar angles $\theta _{lab} > 7 ^{\circ }$, with complete coverage
of azimuthal angles only for $\theta_{lab} > 15 ^{\circ}$.

The overall structure of the detector includes a ``forward'' and a
``barrel'' sector with polar angular coverage from $7\degree -
45\degree$ and $45\degree - 135\degree$, respectively.  Immediately
outside of the beam pipe is an ``inner'' scintillation hodoscope array
segmented in azimuth into 40 (forward) and 20 (barrel) components.
These are used in conjunction with a triple-layer ``outer'' hodoscope
(144 total elements)~\cite{july} to form multiplicity patterns for use
in the trigger.  In the reaction $\pptofourk$ at least three charged
tracks are required within a forward $45\degree$ cone around the beam
axis. The fourth kaon has a maximum polar angle of $60\degree$.  The
inner scintillator array extends to this angle, while the barrel outer
array includes angles up to $135\degree$ to veto events with large
angle tracks.

Between these scintillator arrays are cylindrical drift chambers
(``straws'') which are used for tracking~\cite{straws}.  They are
arranged in 12 planes with alternating horizontal and vertical
orientation in the forward region.  In the barrel, 1400 straws are
formed into a tightly packed bundle aligned with and surrounding the
beam pipe.
  
Three separate devices are incorporated with the purpose of providing
PID information.  These include 3500 silicon-pad counters which
measure energy loss.  They are particularly effective at the low end
of the expected $\beta$ range.  A hodoscope consisting of 24 forward
and 24 barrel threshold \v{C}erenkov counters is positioned just
inside of the outer scintillator array.  These counters are used to
limit the number of ``fast'' particles that are accepted at the
trigger level (typically $\leq 2$) and, for those events which induce
light in the \v{C}erenkov counters, as a measure of the $\beta$ of
that passing particle.  Finally, a RICH detector, having a quartz
radiator ($\beta_{th}=0.64$), was introduced into the forward detector
compartment following the first year of data taking~\cite{RICH}. It is
capable of providing a precise measurement of the $\beta$ for
typically one of the forward-going tracks in each event.  Outside of
all of these components an electromagnetic calorimeter made from lead
and scintillating fibers (Pb/SciFi) is placed.  In the forward region
the shower counter is divided into 300 towers. Each tower is 12.5
radiation lengths thick and has individual phototube
readout~\cite{Calo}.  In the barrel, 24 trapezoidal bars of Pb/SciFi
form a hermetic cylinder with six radiation lengths thickness at
normal incidence.  The barrel calorimeter, in conjunction with the
outer barrel scintillator array, has a special electronic pattern unit
enabling the fast identification of isolated high-energy photons.  The
forward calorimeter is passive, used only in the offline analysis for
evidence of high-energy gammas in the event.
   
The online trigger was based on a charged-particle multiplicity of
four (with no more than one track in the barrel region) coupled with a
crude transverse momentum balance requirement based on the forward
straws multiplicity.  Events with more than two of the threshold
\v{C}erenkov counters responding or with high-energy  gammas in the 
barrel calorimeter were vetoed.  These trigger conditions were tested
using special runs in which such conditions were relaxed in order to
determine the sensitivity of the detector acceptance to the trigger.
For events meeting the trigger conditions, information was recorded
from multiple parallel VME-based subsystems onto a common data
stream~\cite{paulvme} which was recorded on magnetic tape.
Approximately 500 million events were recorded in eight running
periods spanning the time from April 1991 through September 1994.

\section{Data selection}
\subsection{Event reconstruction}

\begin{figure*}
  \begin{center}
    \mbox{\epsfxsize=12.0cm\epsffile{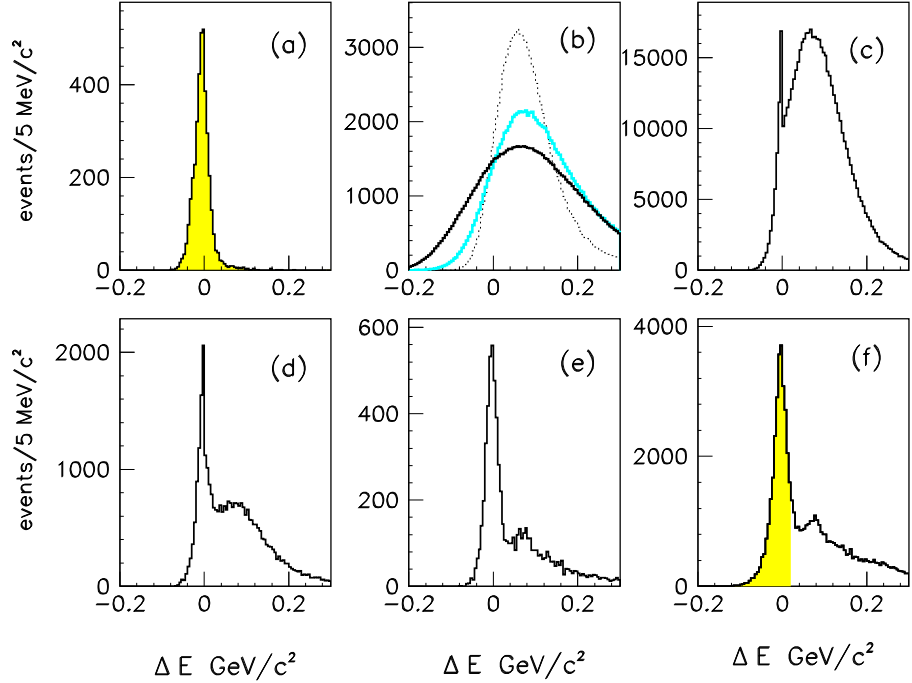}}
  \end{center}

  \caption{ $\Delta E$ (see text) distributions for (a) Monte Carlo
  simulation of the reaction $\pptofourk$ at 1.5 GeV/$c$; (b) Raw
  four-prong data for different momenta ranges: Light line: 1.2
  GeV/$c$, grey line: 1.65 GeV/$c$, black line: 2.0 GeV/$c$; (c)
  Application of the $dE/dx$ criterion only to data in the 1.4-1.45
  GeV/$c$ region; (d) Adding the threshold \v{C}erenkov compatibility
  condition to this data; (e) Adding the criterion from the RICH and
  from the $\gamma$ veto; (f) Final $\Delta E$ distribution for all
  the data taken in the experiment.  The shaded area shows the cut
  used to select the reaction.}

  \label{fig:deltaE}
\end{figure*}

The offline analysis proceeds as follows.  Since the detector is
non-magnetic, only the directions of the four outgoing particles in
the reaction $$\bar p p \to K^+ K^- K^+ K^- \eqno(2)$$ are
measured. The four momenta therefore are obtained by solving the four
equations of energy-momentum conservation.

The conservation of three-momentum from the well-defined initial state
(e.g., $\Delta p/p \approx 10^{-3}$ for the antiproton) allows the
magnitudes of the momenta for the four kaons in the final state to be
expressed as linear functions of one unknown parameter, which can be
taken to be the momentum, $p_4$, of the fourth kaon.  The total energy
in the laboratory of the final state $E=\sum \sqrt{p_i^2 + m_K^2}$ can
then be written as a function of this one parameter $p_4$. The
function $E(p_4)$ has a roughly hyperbolic shape that diverges as $p_4
\to \pm \infty$ and possesses a single minimum. We define $\Delta E =
E(p_4) - E_0$ as the minimum of the function $E(p_4)$ minus the total
energy $E_0$ of the initial state in the laboratory system. This
variable $\Delta E$ is constrained to $\Delta E < 0$ for events from
reaction (1), with a singularity appearing in the available
phase-space at $\Delta E = 0$.  The width of this spike is determined
mainly by the resolution of the straw chamber tracking system which is
well studied.  The resulting resolution on the momentum determination
is 8\% at a typical kaon momentum of 0.5 GeV/$c$.
  
A Monte Carlo simulation of pure $\pptofourk$ events, with the
experimental resolution and all other material effects included,
results in a $\Delta E$ distribution as shown in
Fig.~\ref{fig:deltaE}(a).  The distribution is asymmetric, and its
width increases with increasing incident $\bar p$ momentum.

The four-prong data, prior to application of any PID information, show
a broad $\Delta E$ distribution with no obvious structure.  This is
consistent with expectations since these events are dominated by
background.  The shape variation of such raw $\Delta E$ distributions
over the span of incident momenta explored by our experiment is shown
in Fig.~\ref{fig:deltaE}(b).

\subsection{Particle Identification}

\begin{figure}
  \begin{center}
    \mbox{\epsfxsize=8.6cm\epsffile{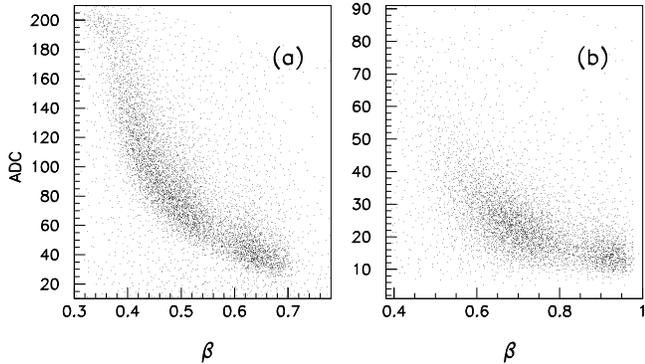}}
  \end{center}

  \caption{ (a) Pulse height in ADC units for the barrel silicon pads
  vs.  $\beta$ for a sample of elastic $\bar p p$ events. (b) Pulse
  height in ADC units for the forward silicon pads vs.  $\beta$ from a
  sample of $p$ or $\bar p$ from the reaction $\bar p p \to \bar p p
  \pi^+ \pi^-$.}

\label{fig:sil}
\end{figure}

In order to isolate the $\pptofourk$ events from the very large
background, use of different selection criteria based on the particle
identification devices is made.  First we require an energy loss
($dE/dx$) compatibility as measured in the silicon detectors.  This is
evaluated by forming a confidence level for each measurement, using a
parametrization of the Landau distribution and integrating the tail
beyond the measured value.  The confidence levels for each measurement
are then combined into a dE/dx confidence level.  The detectors were
calibrated using a sample of well-defined tracks with known momenta.
They come from fully identified $\pbarp$ elastic scattering events and
from $\bar p p \to \bar p p \pi^+ \pi^-$ events. This latter reaction
was isolated almost background free by making use of the
\v{C}erenkov identification and by the requirement of a large energy 
deposit from the outgoing $\bar p$ in the forward electromagnetic
calorimeter~\cite{pap}.  A plot of the pulse height in the silicon
pads vs. $\beta$ is shown in Fig.~\ref{fig:sil}(a) for the barrel
region and in Fig.~\ref{fig:sil}(b) for the forward region. The first
plot makes use of elastic events while the second distribution has
been obtained from $\bar p p \to \bar p p \pi^+ \pi^-$ data.
Application of the $dE/dx$ compatibility requirement on a sample of
the 1.4 to 1.45 GeV/$c$ data results in a $\Delta E$ distribution as
shown in Fig. \ref{fig:deltaE}(c).  At this stage the $\pptofourk$
events are already in evidence.
 
\begin{figure}
  \begin{center}
    \mbox{\epsfxsize=8.6cm\epsffile{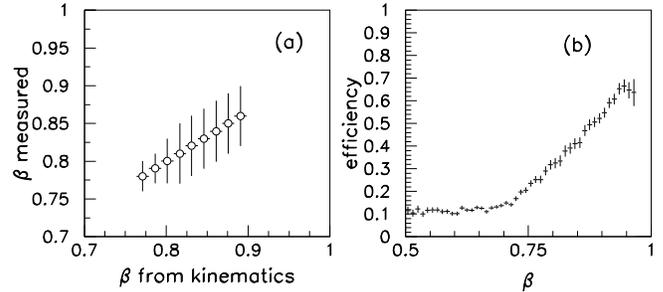}}
  \end{center}

  \caption{ (a) Measured versus expected $\beta$ for elastic events in
  the forward \v{C}erenkov counters with the water radiator.  (b) The
  fraction of tracks giving light in the \v{C}erenkovs as a function
  of $\beta$ for the freon radiator. Events of the type $\pptopppipi$
  are used.  Light observed below threshold results from the
  plexiglass walls of the counters.}

\label{figcer}
\end{figure}

The \v{C}erenkov effect produces a response $R(\beta)$ in
photoelectrons that is monotonically increasing for $\beta \ge
\beta_{th}$.  Two different radiators were used: liquid freon
(FC72)~\cite{fc72}, in the very first data taking only, having a
threshold at $\beta$=0.79, and subsequently water having a threshold
at 0.75.  The result of the $\beta$ measurement from the \v{C}erenkov
response events of the water radiator is compared in
Fig.~\ref{figcer}(a) to the expected $\beta$ from elastic events.

For the freon data, we show in Fig.~\ref{figcer}(b) the ``efficiency''
versus $\beta$ where we have defined efficiency as the ratio between
the $\beta$ distribution of the tracks crossing the \v{C}erenkov
counters which give light to the $\beta$ distribution of all the
tracks.  This study is based on events of the type $\pptopppipi$.  A
smooth increase as a function of $\beta$ is seen which indicates the
expected threshold behavior. The non-zero efficiency below threshold
is caused by light produced in the plexiglass containment wall of the
counters.

The response function of the liquid \v{C}erenkov counters is used to
calculate the expected signal from a passing track of a given
hypothesized momentum.  The expected and measured responses are
compared on the basis of Poisson statistics.  The resulting confidence
levels for the four tracks are then combined with the one obtained
from the $dE/dx$ information to form an overall particle
identification confidence level which is required to be at least 5\%.
The effect of adding the \v{C}erenkov information can be seen in
Fig.~\ref{fig:deltaE}(d).

\begin{figure}
  \begin{center}
    \mbox{\epsfxsize=6.5cm\epsffile{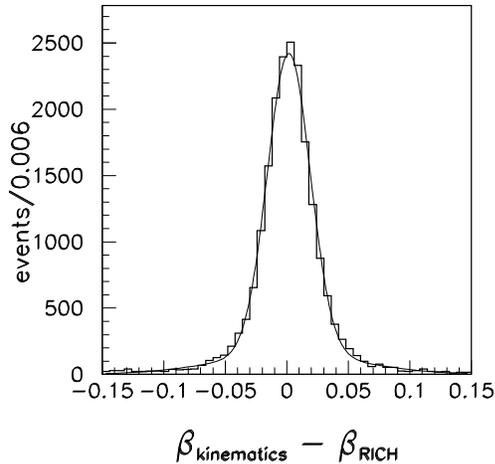}}
  \end{center}

  \caption{ Expected minus measured $\beta$ from the RICH detector for
  elastic events.}

  \label{figrich}
\end{figure}

The response function of the RICH is determined by measurements of
elastic events in which the forward-scattered antiproton of known
momentum penetrates the detector.  Details of this detector
performance are found elsewhere~\cite{RICH}, however a representative
example of the $\beta$ resolution is shown in Fig.~\ref{figrich}.  The
difference between the expected and measured $\beta$ distribution
exhibits a Gaussian behavior with a resolution of $\Delta \beta/\beta
\approx 2 \%$.  In analyzing the response of the detector, we find
that a large fraction of background events produces a RICH $\beta$
measurement which is unphysically greater than 1. The most effective
cut eliminates this background, yet retains the good events, by simply
requiring that the $\beta$ measured by the RICH be less than 0.9.

One background reaction which frequently satisfies the trigger
conditions of the experiment is $$\bar p p \to \bar p p \pi^{+}
\pi^{-} \eqno(3)$$ which is particularly important for $\bar p$
momenta above 1.6 GeV/$c$.  Reaction (3) is identified by the same
selection criteria as those applied to find $\pptofourk$ events.
Events so identified are removed from the final $4K^{\pm}$ sample.

At the final stage of event selection, those events with isolated
$\gamma$'s in the barrel region are removed. This eliminates most of
the events with multiple $\gamma$'s or $\pi^0$'s.  The final $\Delta
E$ distribution after applying all the above selection criteria is
shown in Fig.~\ref{fig:deltaE}(e) for the 1.4 to 1.45 GeV/$c$ sample
and in Fig.~\ref{fig:deltaE}(f) for all of the data taken in the
experiment combined.  A clear signal, peaked at a $\Delta E$ of zero,
is seen which is evidence for events of the type $\pptofourk$.  The
final selection of events is made by requiring $\Delta E \le 20$ MeV.
This cut selects approximately 32 000 events which are retained for
further study.

\section{Acceptance and Luminosity}

The experiment accumulated data at 58 incident $\bar p$ momentum
settings with different luminosities over a period of four years.
However in the following analysis, measurements at the same or nearly
the same momentum have been grouped together in order to reduce
statistical and systematic errors.  The distribution of luminosities
as a function of the momentum are summarized in Table I.

The luminosity was monitored via the known cross section of elastic
$\bar p p$ scattering. Two independent methods were employed to
measure it.  In the first method, the rate of coincidence between
opposite pixels formed in the outer trigger scintillator hodoscope was
used to count elastic events near $90^0$ in the c.m. system. The
second method made use of small microstrip detectors placed near
$90^0$ in the barrel to detect the recoil particle from low-angle
elastic scattering~\cite{strips}. The data have been normalized as a
function of time by using minimum bias scalers which were found to be
very reliable and stable over the lifetime of the experiment. The
relative luminosity error is 2\%.

\begin{figure*}
  \begin{center}
    \mbox{\epsfxsize=14.0cm\epsffile{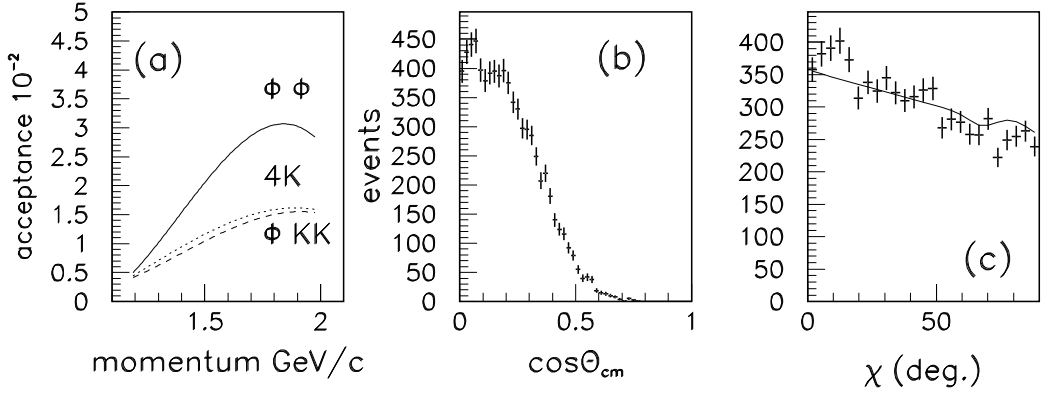}}
  \end{center}

  \caption{ (a) Acceptance for the three reactions $\bar p p \to \phi
  \phi$, $\bar p p \to \phi K^+ K^-$ and $\bar p p \to 4K^{\pm}$.
  Distribution of (b) cos$\Theta_{cm}$ and (c) $\chi$ for the reaction
  $\bar p p \to \phi \phi$ in the momentum range between 1.26 and 1.65
  GeV/c. The data are represented with error bars. In (c) the solid
  line represents the distribution expected for phase space.  }

  \label{fig:two}
\end{figure*}

The acceptance for the elastic and $4K$ events was computed by a GEANT
~\cite{geant} based Monte-Carlo simulation of the physical detector
and trigger conditions.  The Monte-Carlo events were subjected to the
same reconstruction criteria and analysis cuts as the real data.  We
generated Monte-Carlo events for each of the 58 momentum settings,
taking into account the changes in the layout of the apparatus and
trigger conditions. For the physics analysis, three reactions were
generated: $\bar p p \to \phi \phi$, $\bar p p \to \phi K^+ K^-$ and
$\bar p p \to K^+ K^- K^+ K^-$.  The simulation assumed phase space
distributions for all three reactions.  The distributions of the total
acceptance as a function of the incident $\bar p$ momentum are shown
as fitted polynomials in Fig.~\ref{fig:two}(a).

The differential acceptance function for $\bar p p \to \phi \phi$ is
greatly affected by the kinematics of the reaction and the layout of
the detector. The distribution of cos$\Theta_{cm}$ is shown in
Fig. ~\ref{fig:two}(b) for all the data.  Here, $\Theta_{cm}$ is
defined as the centre-of-mass scattering angle of the outgoing $\phi$.
A strong depletion of events occurs in the forward region, due to loss
of tracks in the beam pipe. On the other hand the acceptance in the
$\chi$ angle is almost flat as it can be seen in
Fig. ~\ref{fig:two}(c). Here $\chi$ is defined as the angle formed by
the two $\phi$'s decay planes, in the $\phi \phi$ centre of mass
system. Fig. ~\ref{fig:two}(c) shows the distribution of the $\chi$
angle for the data in the 1.26-1.65 GeV/c region (points with error
bars), together with the distribution of $\chi$ for Monte-Carlo phase
space generated data folded with the acceptance of the apparatus.
This missing forward angle acceptance in $\Theta_{cm}$ affects our
ability to extract total $\phi \phi$ cross sections as described
later.

Systematic errors on the cross section measurements have been
estimated using the known cross section for $\bar p p \to \bar p p
\pi^{+} \pi^{-}$~\cite{pap} whose determination was found to be in
good agreement with previous measurements. We also used the fact that
the cross sections at several energy points were measured several
times under different experimental conditions. We estimate the overall
uncertainty on the absolute scale of the $\phi \phi$ cross section to
be 20~\%.

\section{The data}

\begin{figure}
  \begin{center}
    \mbox{\epsfxsize=8.0cm\epsffile{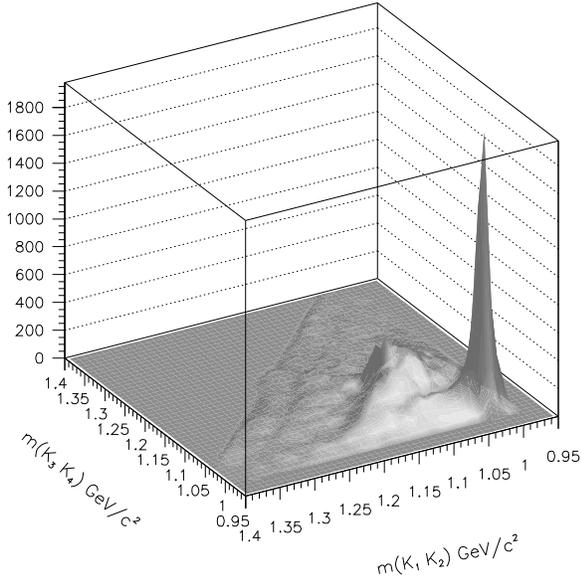}}
  \end{center}

  \caption{ $m(K_3,K_4)$ vs. $m(K_1,K_2)$ (three combinations per
  event) for all the events.  }

  \label{fig:four}
\end{figure}

\begin{figure}[t]
  \begin{center}
    \mbox{\epsfxsize=8.6cm\epsffile{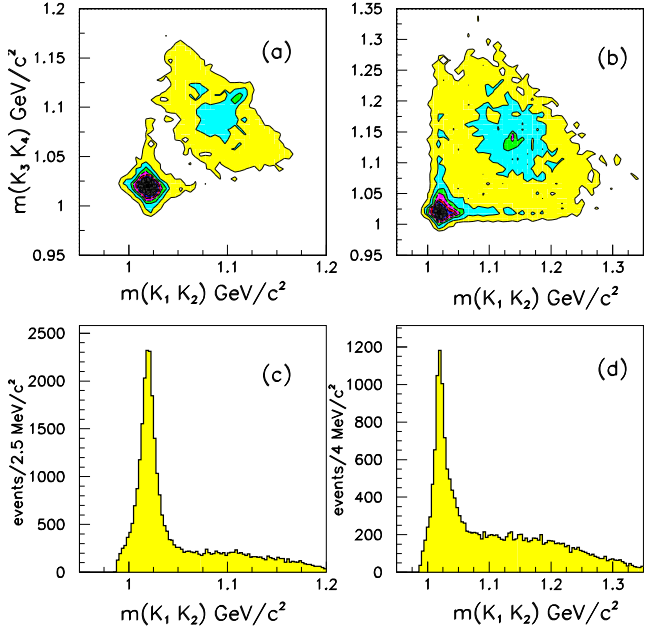}}
  \end{center}

  \caption{ $m(K_3,K_4)$ vs. $ m(K_1,K_2)$ (three combinations per
  event) for a) $\bar p$ incident momentum below 1.5 GeV/$c$ and b)
  above 1.5 GeV/$c$.  c), d) mass projections of a) and b) in the
  $\phi$ band, i.e. plots of $m(K_1 K_2)$ for 1.00 $\le m(K_3,K_4)
  \le$ 1.04 GeV/$c^2$ }

  \label{fig:three}
\end{figure}

For the selected $\bar p p \to K^+ K^- K^+ K^-$ event candidates we
show in Fig.~\ref{fig:four}, in Fig.~\ref{fig:three}(a) and in
Fig.~\ref{fig:three}(b) the scatter plot of the invariant masses
$m(K_3 K_4)$ vs. $ m(K_1 K_2)$.  Fig.~\ref{fig:four} shows all the
available data as a surface plot, where a strong accumulation of
events can be observed at the nominal $\phi \phi$ position.
Fig.~\ref{fig:three}(a) and (b) illustrate the same data under the
form of a scatter diagram for two different regions of incident $\bar
p$ momentum, i.e. below and above 1.5 GeV/$c$ respectively.  Notice
that, due to the absence of charge information, three combinations per
event enter in these plots.
We observe in these distributions a strong enhancement at the position
of the $\phi \phi$. In the higher momentum region we also observe
horizontal and vertical bands which indicate the presence of the $\bar
p p \to \phi K^+ K^-$ final state. Monte-Carlo simulations confirm
that the diagonal bands represent the reflection of the $\phi \phi$
peak due to the multiple combinations.  Selecting one $\phi$,
i.e. requiring one $m(K_3 K_4)$ combination to lie in the region
1.00--1.04 GeV/$c^2$ and plotting the opposite combination $m(K_1
K_2)$, we obtain the distributions shown in Fig. ~\ref{fig:three}(c)
and (d) where a clean $\phi$ peak is visible. The structure is well
centered at the nominal $\phi$ mass.

\section{Channel Likelihood fit}

In order to separate the $\phi \phi$ cross section from the $\phi
K^+K^-$ final state and from the $4K$ and other background mixture,
the channel likelihood technique~\cite{chan} was used.  The method
performs a maximum likelihood fit to the data using three amplitudes:
$\phi \phi$, $\phi K^+K^-$ and phase space (which is a mixture of $4K$
and background and cannot be separated at this stage).  The different
channels have been described by the following amplitudes:
\[
\phi \phi : A_{\phi \phi} = \sum_{i,j=1}^{3} B_i(m_{KK})\times B_j(m_{KK}),
\]
\[
\phi KK : A_{\phi KK} = \sum_{i=1}^{6} B_i(m_{KK}).
\]
The likelihood function employed is the following:
\[
{\mathcal{L}} = x_{\phi \phi} \frac{A_{\phi \phi}}{I_{\phi \phi}} +
x_{\phi KK} \frac{A_{\phi KK}}{I_{\phi KK}} + (1 - x_{\phi \phi} -
x_{\phi KK}).
\]
In the above expressions $x_{\phi \phi}$ and $x_{\phi KK}$ represent
the fractions of the $\phi \phi$ and $\phi K^+K^-$ channels
respectively, $B_i(m_{KK})$ represents the $\phi$ lineshape functions,
$I_{\phi \phi}$ and $I_{\phi KK}$ represent the corresponding
normalization integrals for the two amplitudes describing the two
channels respectively.

The $\phi$ width is a narrow resonance ($\Gamma$=4.4 MeV) therefore
the observed width in this experiment is dominated by the detector
resolution which we found to be non-Gaussian.  In order to describe
the $\phi$ lineshape we made use of the Monte-Carlo simulations. The
best representation of this lineshape is obtained by means of a
relativistic spin 0 Breit-Wigner form whose full width varies from 11
to 15 MeV as the incident $\bar p$ momentum increases from 1.1 to 2.0
GeV/c.

\begin{figure}[t]
  \begin{center}
    \mbox{\epsfxsize=8.6cm\epsffile{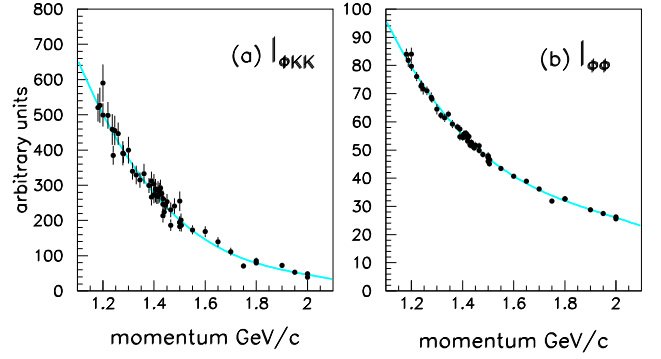}}
  \end{center}

  \caption{ Variation of the normalization integrals for the
  amplitudes describing the $\phi K^+K^-$ (a) and $\phi \phi$ (b)
  amplitudes as functions of the incident $\bar p$ momentum.}

  \label{fig:five}
\end{figure}

The normalization of the amplitudes was computed numerically using the
Monte-Carlo simulations of the reaction $\bar p p \to 4K$. The
resulting integrals were then fit to polynomials as functions of the
incident $\bar p$ momentum. Changes in the layout of the apparatus and
trigger conditions have little effect on the values of these integrals
as seen in Fig.~\ref{fig:five} where these integrals and the fitted
polynomials are shown.

\begin{figure}
  \begin{center}
    \mbox{\epsfxsize=8.6cm\epsffile{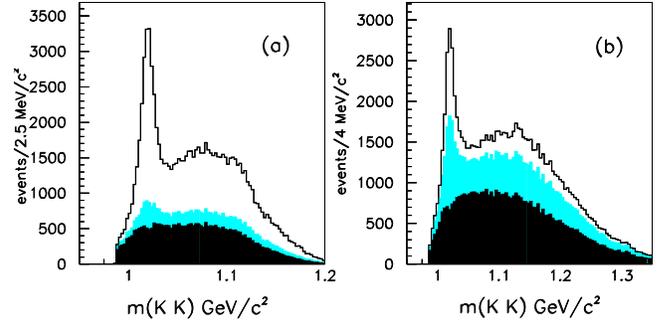}}
  \end{center}

  \caption{ Combinatorial m(KK) distribution (six entries per event)
  for incident $\bar p$ momentum (a) below and (b) above 1.5 GeV/c.
  The white region represents the $\phi \phi$ contribution, the grey
  region shows the $\phi K^+ K^-$ contribution and the black region
  shows the phase space (background + $4K$) contribution.  }

  \label{fig:six}
\end{figure}

The channel likelihood fit gives the fractions of events as well as a
probability for each event to belong to one of the three
hypotheses. The results of the fits are visually displayed in
Fig.~\ref{fig:six}. Here we show the combinatorial $m(KK)$
distribution (six entries per event) for incident $\bar p$ momentum
below and above 1.5 GeV/c. The white region represent the $\phi \phi$
contribution, the grey region shows the $\phi K^+ K^-$ contribution
while the black region shows the phase space (background + $4K$)
contribution. We notice that no $\phi$ peak has been left in the phase
space distribution, indicative of a successful fit.  We also notice
the strong increase of the $\phi K^+K^-$ contribution in the higher
momentum region. The analysis gives a total of approximately 11 400
$\phi \phi$ events.

\begin{figure*}
  \begin{center}
    \mbox{\epsfxsize=14.0cm\epsffile{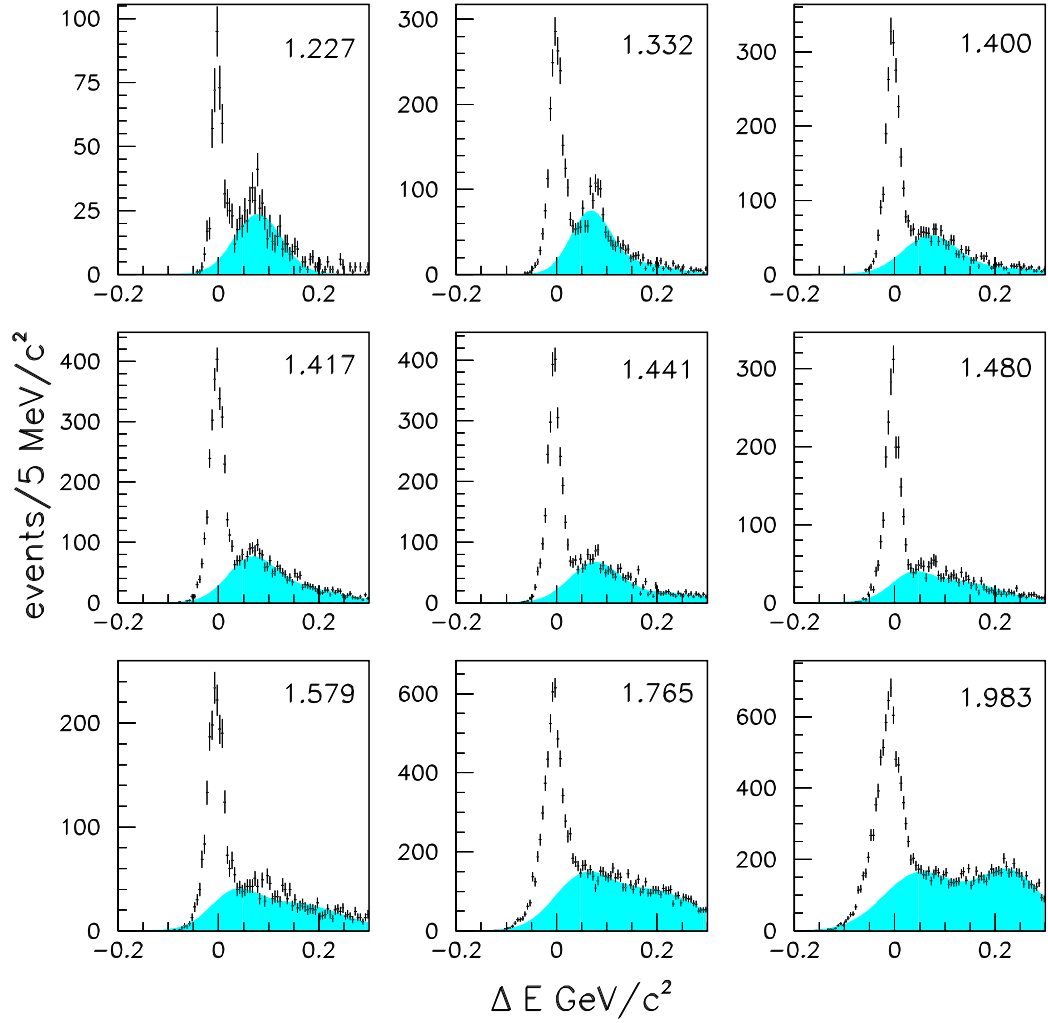}}
  \end{center}

  \caption{ $\Delta$ E distributions for the nine regions in
  increasing values of incident $\bar p$ momentum.  The grey area
  represents the estimated background.  }

  \label{fig:seven}
\end{figure*}

The channel likelihood method is able to separate the three $\phi
\phi$, $\phi K^+K^-$ and (background + $4K$) contributions. In order
to separate the $4K$ contribution from the background, the $\Delta E$
distributions as discussed in section III.A.  For this purpose, in
order to reduce the errors, a further compression of the data was
performed, grouping them in nine slices of incident $\bar p$ momentum.
The $\Delta E$ distributions for all intervals are shown in
Fig.~\ref{fig:seven}.  A clean peak at $\Delta E = 0$ over some
background is observed which represents the total amount of the
reaction $\bar p p \to 4K$ including $\phi \phi$ and $\phi K^+K^-$
contributions.  These distributions have been fitted using a
Monte-Carlo generated shape for the $\Delta E$ distribution which
includes simulations at all energies of $\phi \phi$ and $4K$ final
states, and a background parametrization using the sum of two
Gaussians.  The number of non-resonant $4K$ events was therefore
obtained by
\[
N_{4K} = N_T - N_{\phi \phi} - N_{\phi KK} - N_{b}
\]
In the above expression, the number of $4K$ events is the difference
between the total number of events in each bin ($N_T$), the $\phi
\phi$ ($N_{\phi \phi}$) and $\phi K^+K^-$ ($N_{\phi KK}$) yields which
were obtained from the channel likelihood fits, and the background
($N_{b}$) is drawn from the fits to the $\Delta E$ distributions. Due
to the uncertainty in the background subtraction, a 50\% systematic
error has been added quadratically to the statistical errors.  The
background below the $4K$ signal is relatively small, being in average
of the order of 10 \% increasing to 20 \% only in the higher momentum
regions.

\section{Cross sections}

\begin{figure}
  \begin{center}
    \mbox{\epsfxsize=8.6cm\epsffile{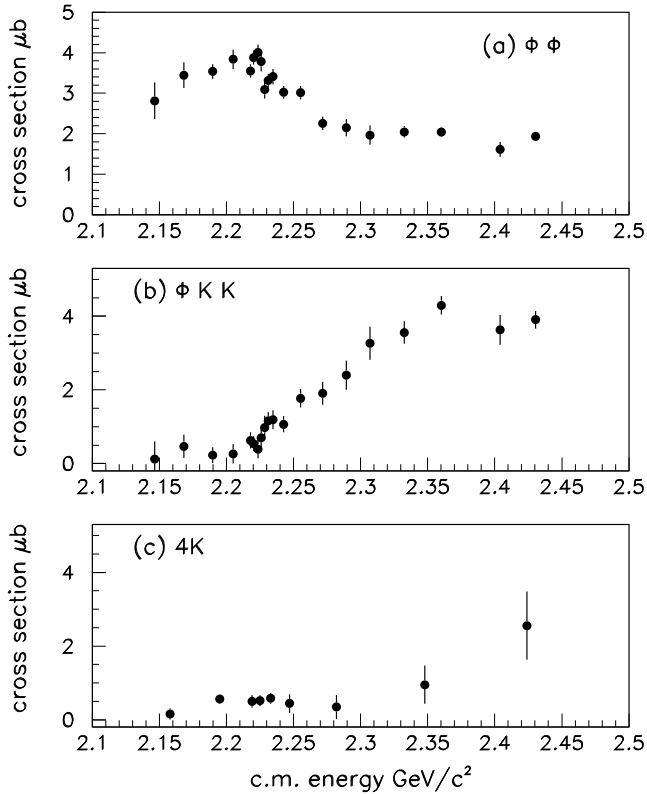}}
  \end{center}

  \caption{ Cross sections in $\mu$b for the reactions (a) $\bar p p
  \to \phi \phi$, (b) $\bar p p \to \phi K^+K^-$ and (c) $\bar p p \to
  2K^+ 2K^-$ corrected for unseen $\phi$ decay modes.}

  \label{fig:eight}
\end{figure}

Having determined the number of events for each channel we have
computed the corresponding cross sections as: $$\sigma =
\frac{Events}{Acceptance \times Luminosity}$$ Due to decreasing
performance of the threshold \v{C}erenkov counters in the last period
of the data taking, part of data suffer an additional 20\% systematic
normalization error.  These data represent about 30\% of the total and
have not been used in the calculation of the cross sections. However,
properly scaled, these data can be used in the study of the angular
distributions.  These cross sections have been corrected for unseen
$\phi$ decay modes and are displayed in Table I and shown in
Figs. \ref{fig:eight}.

\begin{table*}
  \caption {Cross sections for $\bar p p \to \phi \phi$, $\phi K^+K^-$
  and $4K$}
  \label{tab1} 
  \begin{tabular}{ccccccc} 
Momentum region & c.m. &Luminosity& $\sigma(\phi \phi)$ &
$\sigma(\phi K^+ K^-$) &  c.m. & $\sigma(4 K^{\pm})$ \\
 & Energy & & & & Energy & \\
GeV/$c$ & GeV/$c^2$ & nb$^{-1}$ & $\mu$b & $\mu$b & GeV/$c^2$ & $\mu$b \\ 
  \hline
1.188-1.200 & 2.147 & 26.8 & 2.86 $\pm$ 0.46 & 0.25 $\pm$ 0.49 & 
2.158 & 0.15 $\pm$ 0.15 \\ 
1.237-1.278 & 2.168 & 32.8 & 3.45 $\pm$ 0.31 & 0.96 $\pm$ 0.32 & & \\
1.300-1.330 & 2.190 & 69.1 & 3.54 $\pm$ 0.18 & 0.23 $\pm$ 0.21 &
2.195 & 0.56 $\pm$ 0.13 \\
1.360 & 2.205 & 33.4 & 3.84 $\pm$ 0.24 & 0.27 $\pm$ 0.26 & & \\
1.390-1.400 & 2.218 & 60.7 & 3.55 $\pm$ 0.17 & 0.63 $\pm$ 0.22 &
2.219 & 0.50 $\pm$ 0.17 \\
1.404-1.405 & 2.220 & 53.9 & 3.88 $\pm$ 0.17 & 0.54 $\pm$ 0.21 & & \\
1.410-1.415 & 2.223 & 38.0 & 4.00 $\pm$ 0.20 & 0.39 $\pm$ 0.25 &
2.225 & 0.52 $\pm$ 0.15 \\
1.420 & 2.226 & 25.7 & 3.84 $\pm$ 0.24 & 0.71 $\pm$ 0.29 & & \\
1.425-1.430 & 2.228 & 26.4 & 3.09 $\pm$ 0.22 & 0.97 $\pm$ 0.32 & & \\
1.435 & 2.231 & 47.7 & 3.33 $\pm$ 0.17 & 1.16 $\pm$ 0.235&
2.233 & 0.58 $\pm$ 0.15  \\
1.440-1.450 & 2.235 & 35.1 & 3.41 $\pm$ 0.19 & 1.19 $\pm$ 0.26 & &\\
1.465-1.480 & 2.242 & 48.7 & 3.02 $\pm$ 0.15 & 1.07 $\pm$ 0.22 &
2.247 & 0.44 $\pm$ 0.26 \\
1.500-1.506 & 2.255 & 42.7 & 3.01 $\pm$ 0.16 & 1.77 $\pm$ 0.25 & &\\
1.550 & 2.272 & 24.3 & 2.25 $\pm$ 0.17 & 1.90 $\pm$ 0.31 &
2.282 & 0.34 $\pm$ 0.33 \\
1.600 & 2.289 & 20.1 & 2.15 $\pm$ 0.21 & 2.40 $\pm$ 0.40 & &\\
1.650 & 2.307 & 13.6 & 1.97 $\pm$ 0.24 & 3.26 $\pm$ 0.45 & & \\
1.700-1.750 & 2.333 & 32.6 & 2.04 $\pm$ 0.14 & 3.56 $\pm$ 0.30 &
2.348 & 0.95 $\pm$ 0.52 \\
1.800 & 2.360 & 39.3 & 2.04 $\pm$ 0.12 & 4.30 $\pm$ 0.25 & & \\
1.900-1.950 & 2.404 & 15.4 & 1.61 $\pm$ 0.18 & 3.63 $\pm$ 0.41 &
2.424 & 2.55 $\pm$ 0.93 \\
2.000 & 2.430 & 52.1 & 1.93 $\pm$ 0.11 & 3.91 $\pm$ 0.24 & & \\
\end{tabular}
\end{table*}

The $\phi \phi$ cross section has been corrected assuming phase space
in the calculation of the acceptance. This is not really a strong
assumption as it can be seen from Fig.~\ref{fig:two}(c).  In addition,
a spin parity analysis of the $\phi \phi$ final state has been
performed~\cite{pwa}. This analysis shows that the $\phi \phi$ system
is dominated by $J^{PC}=2^{++}$. Correcting the mass spectrum with the
results from the spin-parity analysis has little influence on the
shape of the integrated acceptance as a function of the $\phi \phi$
mass.

Notice that:
\begin{itemize}
\item  
The $\phi \phi$, $\phi K^+K^-$ and $4K^{\pm}$ cross sections have
different shapes.  The $\phi \phi$ cross section, in particular, has a
strong threshold enhancement while the $\phi K^+K^-$ and $4K^{\pm}$
cross sections have a smooth increase as a function of the centre of
mass energy.
\item 
The $\phi \phi$ cross section is rather large, about $3.5 \mu b$ in
the threshold region.
\item
No evidence for narrow structures is found.
\end{itemize}

The large $\phi \phi$ production close to threshold can be interpreted
as a violation of the OZI rule. If the OZI rule is interpreted to
forbid strangeness production in $\bar p p$ annihilations then the
process $\bar p p \to \phi \phi$ can proceed only via the small $\bar
u u, \bar d d$ component present in the $\phi$ wave function. With a
deviation from ideal mixing $\theta - \theta_0$ of only $1^0$ to
$4^0$, the $\phi$ is nearly 100~\% $\bar s s$.  We can therefore
derive an upper limit tan$^4(\theta-\theta_0)\approx 2.5\times
10^{-5}$ for the ratio of cross sections $\sigma_{\phi
\phi}/\sigma_{\omega \omega}$ for production in $\bar p p$
annihilation. Although the cross section of $\bar p p \to \omega
\omega$ has not been measured directly, an estimate can be obtained
from the total $2\pi^+ 2\pi^- 2\pi^0$ cross section~\cite{omom}, which
was measured to be about 5 mb in the energy range of our experiment.
There are many reaction channels that contribute to this final state.
If we estimate that it is 10 \% $\omega \omega$ then the expected
$\phi \phi$ cross section is 10 nb, two orders of magnitude lower than
our measurements.

\section{Search for $\xi/{\lowercase {f}}_J(2230)$}

\begin{figure}
  \begin{center}
    \mbox{\epsfxsize=8.6cm\epsffile{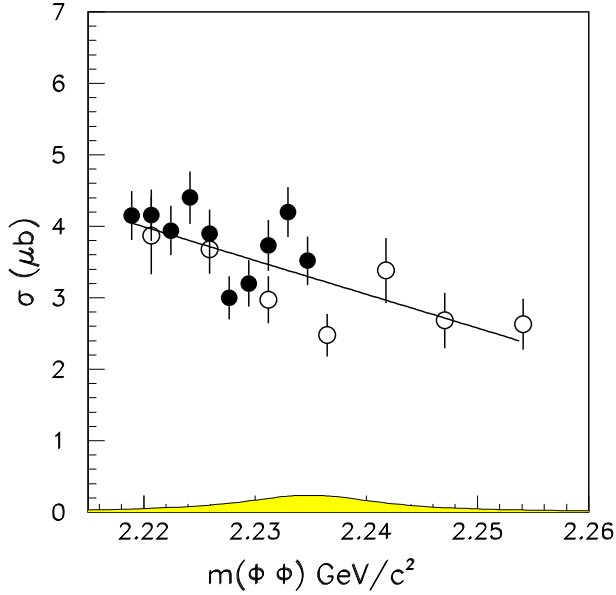}}
  \end{center}

  \caption{ Cross section in $\mu$b for the reaction $\bar p p \to
  \phi \phi$ in a fine scan over two different periods of data
  collection corrected for unseen $\phi$ decay modes.  Open circles:
  1991 data; full circles: 1993 data.  The line is the result from the
  fit described in the text, the curve represents a Breit-Wigner
  resonance whose amplitude is at the 95 \% c.l. upper limit for the
  production of a $\xi/f_J(2230)$ with a mass of 2235 MeV and a width
  of 15 MeV.}

  \label{fig:nine}
\end{figure}

To search for the $\xi/f_J(2230)$ resonance, a fine scan of the $\phi
\phi$ cross section was performed during two different data taking
periods.  The results from these scans are shown in
Fig.~\ref{fig:nine} and displayed in Table II.

\begin{table}
\caption {$\phi \phi$ cross section for the fine scan.}
\label{tab2} 
\begin{tabular}{cc|cc} 
{c.m. energy} &{$\sigma(\phi \phi)$} & {c.m. energy} &{$\sigma(\phi \phi)$} \\
GeV/$c^2$ & {$\mu b$} & GeV/$c^2$ & {$\mu b$}\\ 
\hline
2.219 & 4.15  $\pm$ 0.34 & 2.231 & 3.73  $\pm$ 0.35 \\ 
2.221 & 4.16  $\pm$ 0.36 & 2.231 & 2.97  $\pm$ 0.33 \\
2.221 & 3.87  $\pm$ 0.54 & 2.233 & 4.20  $\pm$ 0.35 \\
2.222 & 3.94  $\pm$ 0.35 & 2.235 & 3.52  $\pm$ 0.34 \\
2.224 & 4.40  $\pm$ 0.36 & 2.236 & 2.48  $\pm$ 0.30 \\
2.226 & 3.90  $\pm$ 0.34 & 2.242 & 3.38  $\pm$ 0.45 \\
2.226 & 3.68  $\pm$ 0.34 & 2.247 & 2.68  $\pm$ 0.39 \\
2.228 & 3.00  $\pm$ 0.30 & 2.254 & 2.63  $\pm$ 0.35 \\
2.229 & 3.20  $\pm$ 0.32 \\
\end{tabular}
\end{table}

Here the empty and full dots distinguish the two different sets of the
data, showing good agreement in the size of the $\phi \phi$ cross
section. No narrow structure is visible in the data.  We have fitted
the $\phi \phi$ mass spectrum using a polynomial and a Breit-Wigner
form representing the $\xi/f_J(2230)$ with m=2235 MeV and $\Gamma$=15
MeV, parameters measured in the reaction $J/\psi \to \gamma \xi$ where
$\xi \to \bar p p$ by the BES experiment~\cite{bes}.  The Breit-Wigner
form is written as:
\begin{eqnarray*}
\sigma_{BW} = & (w_{i}w_{f}) \times
                \frac{(2J+1)}{(2S_{1}+1)(2S_{2}+1)} \times
                \frac{4\pi(\hbar c)^{2}}{s-4m_{p}^{2}} \\
              & \times \frac{\Gamma^{2}}{(\sqrt{s}-m_{res})^{2} + \Gamma^{2}/4}.
\end{eqnarray*}
Here $(w_{i}w_{f})$ is the double branching ratio, $w_{i}w_{f} =
\rm{BR}(X\rightarrow\pbarp)\times\rm{BR}(X\rightarrow\phi\phi)$\@. The
$S_{i}$ terms are the spins of the initial proton and antiproton
(1/2), and $J$ is the total angular momentum of the resonance,
reducing the angular momentum term, $(2J+1)/ \left(
\left(2S_{1}+1)(2S_{2}+1\right)\right)$ to $5/4$ in the case of a
$J = 2$ resonance.  A limit on the product of the branching ratios of:
$\rm{BR}(\xi\rightarrow\pbarp)\times\rm{BR}(\xi\rightarrow\phi\phi)
\le 6 \times 10^{-5}$ at 95 \% c.l. is obtained.

\section{Conclusions}

We have performed a high statistics study of the reaction $\bar p p
\to 4K^{\pm}$ using in-flight $\bar p$ from 1.1 to 2.0 GeV/c incident
momentum interacting on a hydrogen jet target of the JETSET (PS202)
experiment at CERN/LEAR.  The reaction is dominated by a strong $\phi
\phi$ production at threshold whose strength exceeds by two orders of
magnitude the yield extracted from a simple interpretation of the OZI
rule.
 
Several models have been proposed in order to explain the large OZI
violations observed in hadron induced reactions and particularly in
some $\bar p p$ annihilation channels. However, few quantitative
calculations exist for the specific channel under investigation in the
present experiment.

A model which interprets $\bar p p$ annihilations to $\phi \phi$ as
due to $K \bar K$ rescattering~\cite{locher} is able to predict the
order of magnitude of the cross section ($\approx 2.4 \mu b$), but not
the detailed shape of the observed spectrum.  Other models make use of
hyperon-antihyperon intermediate states~\cite{mull}.  In this case the
size of the cross section is underestimated by a factor of about 4.
Other ways to enhance production of the $\phi \phi$ system have been
suggested in ref.~\cite{ellis} by invoking the hypothesis of intrinsic
strangeness content in the proton.  The authors suggest that $\phi
\phi$ production could originate from rearrangement diagrams with
strange quarks originating from the proton sea which are polarized
with total spin S=1.  The authors conclude that these connected
rearrangement diagrams very likely mask any possible glueball
resonance contributions which are expected to be dominant among the
disconnected diagrams.  Further information and possible new inputs to
the problem may come from a spin analysis of the observed $\phi \phi$
threshold enhancement ~\cite{pwa}.

No evidence for narrow resonance is found and we set an upper limit
for the production of $\xi/f_J(2230)$ with m=2235 MeV and $\Gamma$=15
MeV of:
$\rm{BR}(\xi\rightarrow\pbarp)\times\rm{BR}(\xi\rightarrow\phi\phi)
\le 6 \times 10^{-5}$ at 95 \% c.l. Combined with our scan of $\bar p
p \to K^0_S K^0_S$~\cite{jetset-2} a consistent and stringent
rejection of a narrow resonance ($\Gamma$ $<$ 30 MeV) appears.

\section{Acknowledgments}
We thank the teams of the CERN Antiproton Complex, in particular the
LEAR staff. This work has been supported in part by CERN, the German
Bundesministerium f\"{u}r Bildung, Wissenschaft, Forschung und
Technologie, the Italian Istituto Nazionale di Fisica Nucleare, the
Swedish Natural Science Research Council, the Norwegian Research
Council, and the United States National Science Foundation.


\begin{references}
\bibitem{fc72}
FC72 is the trade name of a compound marked by 3M composed of $C_6 F_{14}$,
sometimes known as ``liquid freon''.
\end{references}
\end{document}